\documentclass[aps,preprint]{revtex4}%
\usepackage{amsfonts}
\usepackage{amsmath}
\usepackage{amssymb}
\usepackage{graphicx}%
\providecommand{\U}[1]{\protect\rule{.1in}{.1in}}
\begin{document}
\preprint{ }
\title{Search For Gravitational Waves Through the Electromagnetic Faraday Rotation}
\author{M. Halilsoy}
\email{mustafa.halilsoy@emu.edu.tr}
\author{O. Gurtug}
\email{ozay.gurtug@emu.edu.tr}
\affiliation{Department of Physics, Eastern Mediterranean University, G.Magusa, North
Cyprus, Mersin 10 - Turkey}
\pacs{PACS number}

\begin{abstract}
A method is given which renders indirect detection of strong gravitational
waves possible. This is based on the reflection (collision) of a linearly
polarized electromagnetic shock wave from (with) a cross polarized impulsive
and shock gravitational waves in accordance with the general theory of
relativity. This highly non-linear process induces a detectable Faraday
rotation in the polarization vector of the electromagnetic field.

\end{abstract}
\maketitle

\section{INTRODUCTION}

Nonlinear interaction (collision) of gravitational waves (GWs) in Einstein's
theory, which was initiated long time ago by the two Letters to Nature
\cite{KP}\cite{SP} is a well known subject by now \cite{GJB}\cite{BH}. As an
outcome of this process we quest whether it is possible to probe controllable
electromagnetic (em) waves in sensing the passing of a strong GW. Given the
known exact solutions to Einstein's field equations to date the answer is
affirmative. After encounter with a GW, the physical changes in the em wave
will inform about the latter. We recall that the first indirect evidence of
GWs from the binary pulsar in the early 1970s, by the team of R. Hulse and J.
Taylor also was based on em observations. The main feature on which we
concentrate here is the rotation of em polarization vector known as the
Faraday effect. Such rotation has been introduced also for the GWs
\cite{TA}\cite{ASM}\cite{WNG}\cite{PS}, however its detection is hampered by
the challenge of detecting GWs themselves.

In this paper, we propose that given its Faraday rotation after encountering
with a GW, a linearly polarized em wave signals the passage of a GW. For a
thorough analysis GWs may be considered in three different profiles. The first
( and simplest), albeit idealized, is an impulsive GW whose possible sources
are cosmic bursts, supernova explosions, collision of black holes, etc. It is
dubbed "idealized" in the sense that it arises as a limit of a Gaussian
curvature. The second profile of interest is shock type with a uniform
curvature. The third type is a hybrid sandwich GW that lies intermediate to
the aforementioned ones. In the sequel, we shall restrict ourselves to the two
types alone.

The general form of plane waves known as the Rosen form is described by the
line element%

\begin{equation}
ds^{2}=2e^{-M}dudv-e^{-U}\left\{  \left[  e^{V}dx^{2}+e^{-V}dy^{2}\right]
\cosh W-2\sinh Wdxdy\right\}  .
\end{equation}

The metric functions depend only on the null coordinates $u$ and $v$. In the
plane wave regions, all metric functions appeared in this line element depend
only on either $u$ or $v$ . In this paper we \ shall be interested in the
metric function $W(u,v)$ which represents the second polarization content of
the waves.

\section{INTERACTION OF ELECTROMAGNETIC AND GRAVITATIONAL WAVES}

In our analysis, we consider the interaction of linearly polarized plane em
waves with a cross polarized GWs that propagate in the opposite directions in
each of the incoming regions as illustrated in Fig.1. Region III $(v>0,u<0)$,
contains linearly polarized plane em wave described by the line element%

\begin{equation}
ds^{2}=2dudv-\left(  \frac{1}{2}+g\right)  \left(  dx^{2}+dy^{2}\right)  ,
\end{equation}

where $g=g(v)$ is only a function of $v$. The only non - trivial
Einstein-Maxwell equation in this region is given by%

\begin{equation}
2U_{vv}-U_{v}^{2}=4\Phi_{00},
\end{equation}

where $e^{-U}=\frac{1}{2}+g$ and $\Phi_{00}$ stands for the Ricci tensor.

Region II $(v<0,u>0)$, contains plane GW described by%

\begin{equation}
ds^{2}=2dudv-\left(  \frac{1}{2}+f\right)  \left(  Zd\overline{x}^{2}%
+Z^{-1}d\overline{y}^{2}\right)  ,
\end{equation}

in which$\ \ f$ and $Z=e^{V}$ depend only on $u$ and are constrained by the
vacuum Einstein equation%

\begin{equation}
2U_{uu}-U_{u}^{2}=V_{u}^{2}.
\end{equation}

Now, we rotate the $\left(  \overline{x},\overline{y}\right)  $ axes by angle
$\alpha$ to align them along with $\left(  x,y\right)  $ at the cost of
creating a cross polarization term in (4). The two waves, em from left and GW
from right make a head on collision on the hypersurface specified by $u=0=v$.
Physically the cross $(\times)$ mode of the GW will rotate the linear $(+)$
mode of the em wave giving rise to the Faraday effect in the em wave.
Irrespective of the initial data the Einstein - Maxwell equations admit the
following metric and Maxwell fields in the interaction region $\left(
u>0,v>0\right)  $ \cite{GJB};%

\begin{align}
ds^{2}  &  =\left(  \frac{-g^{^{\prime}}\sqrt{\frac{1}{2}+f}}{\sqrt{f+g}%
\sqrt{\frac{1}{2}-g}}\right)  dudv-\left(  f+g\right)  \left[  \left(
e^{V}dx^{2}+e^{-V}dy^{2}\right)  \cosh W-2\sinh Wdxdy\right]  ,\\
\Phi_{0}  &  =\left(  \frac{-g^{^{\prime}}}{2\sqrt{f+g}\sqrt{\frac{1}{2}-g}%
}\right)  e^{i\beta(u)},\nonumber\\
\Phi_{2}  &  =-\frac{\sqrt{\frac{1}{2}-g}}{2\sqrt{f+g}}\left(  V^{^{\prime}%
}\cosh W+iW^{^{\prime}}\right)  e^{i\beta(u)},\nonumber
\end{align}

in which the phase function $\beta(u)$ is to be determined from $\beta
^{^{\prime}}=-\frac{1}{2}V^{^{\prime}}\sinh W$ ( a prime implies derivative
with respect to the argument for each function). The incoming em shock wave
with $(+)$ mode is characterized by the function%

\begin{equation}
g(v)=\frac{1}{2}-\sin^{2}\left(  bv_{+}\right)  ,
\end{equation}

where $b$ is the energy (frequency) constant and $v_{+}=v\theta(v)$ with
$\theta(v)$ the unit step function. The initial data for the GW are;%

\begin{align}
\text{i) \ \ for\ impulsive wave, \ \ \ }f  &  =\frac{1}{2}-u_{+}^{2},\text{
\ \ \ \ \ \ \ }Z=\frac{1-u_{+}}{1+u_{+}},\\
\text{ii) \ \ \ for\ \ shock wave, \ \ \ \ \ }f  &  =-\frac{1}{2}+\cos
u_{+}\cdot\cosh u_{+},\text{ \ \ \ \ \ \ }Z=\frac{\cos u_{+}}{\cosh u_{+}%
},\nonumber
\end{align}

where $u_{+}$ is to be understood with the step function, i.e, $u\theta(u).$
The resulting solution for both cases is summarized as follows;%

\begin{align}
e^{2V}  &  =\frac{Z^{2}\cos^{2}\left(  \alpha/2\right)  +\sin^{2}\left(
\alpha/2\right)  }{Z^{2}\sin^{2}\left(  \alpha/2\right)  +\cos^{2}\left(
\alpha/2\right)  },\\
\sinh W  &  =\frac{1}{2}\left(  Z-Z^{-1}\right)  \sin\alpha,\nonumber\\
\tan\beta &  =\frac{1}{2Z\cos\alpha}\left[  \sqrt{4Z^{2}+\left(
1-Z^{2}\right)  ^{2}\sin^{2}\alpha}-\left(  1+Z^{2}\right)  \sin\alpha\right]
,\nonumber
\end{align}

while $f(u)$ and $g(v)$ functions are readily available from (7) and (8). We
note that $\alpha$ is confined by $0<\alpha<\pi/2.$ Before collision the em
wave is linearly polarized along the $x$-axis.

\section{THE FARADAY ROTATION}

The polarization vector of the incoming em wave which was aligned with the
$x$-axis does not preserve its linear form after interacting with the
gravitatinal wave. In an orthonormal tetrad $\left(  \omega^{a}\right)  $, the
line element can be expressed by%

\begin{equation}
ds^{2}=\left(  \omega^{0}\right)  ^{2}-\left(  \omega^{1}\right)  ^{2}-\left(
\omega^{2}\right)  ^{2}-\left(  \omega^{3}\right)  ^{2}.
\end{equation}

The electric and magnetic field components are defined by \cite{HM}%

\begin{align}
E_{x}  &  =F_{02}=\operatorname{Re}\left(  \Phi_{0}-\Phi_{2}\right)  ,\\
H_{y}  &  =F_{12}=-\operatorname{Re}\left(  \Phi_{0}+\Phi_{2}\right)
,\nonumber\\
E_{y}  &  =F_{03}=\operatorname{Im}\left(  \Phi_{0}+\Phi_{2}\right)
,\nonumber\\
H_{x}  &  =F_{31}=\operatorname{Im}\left(  \Phi_{0}-\Phi_{2}\right)
.\nonumber
\end{align}

The Faraday rotation angle $\theta$ is determined from the electric field
components of the em wave by%

\begin{equation}
\tan\theta=\frac{\operatorname{Im}\left(  \Phi_{0}+\Phi_{2}\right)
}{\operatorname{Re}\left(  \Phi_{0}-\Phi_{2}\right)  },
\end{equation}

in which $\Phi_{0}$ and $\Phi_{2}$ are the em spinor components given in Eq.
(6). In terms of metric functions, the Faraday rotation angle is

\begin{equation}
\tan\theta=\frac{W^{^{\prime}}+\left(  V^{^{\prime}}\cosh W-2b\cot bv\right)
\tan\beta}{W^{\prime}\tan\beta-V^{^{\prime}}\cosh W-2b\cot bv}.
\end{equation}

In the case of gravitational plane impulsive wave and for the particular angle
$\alpha=45^{\circ}$, Eq.(13) becomes%

\begin{equation}
\tan\theta=\left(  1-u^{2}\right)  \left\{  \frac{1-A(u)b\cot bv}%
{A(u)+b\left(  1-u^{2}\right)  ^{2}\cot bv}\right\}  ,
\end{equation}

where $A(u)=1+u^{2}-\sqrt{2}\sqrt{1+u^{4}}.$ The behavior of this expression
for various $b$ values is illustrated in Figure 2. The behavior for shock
gravitational wave case which can be obtained \ in anology to Eq. (14) also
from the Eqs. (8-13) is given in Figure 3. Since the $v$ dependence in
$\tan\theta$ is periodic, one would expect to see this behavior explicitly.
However, due to the curvature singularity that the spacetime possesses on the
hypersurface, $u^{2}+\sin^{2}bv=1$, we have to choose $0<u<1$ and $0<bv<\pi/2$
such that $u^{2}+\sin^{2}bv<1,$ for all ($u,v$). As a result of this
constraint condition the $v$ dependence does not display its periodic
character in this range. Figures 2a, 2b and 2c represent the impulsive wave
plots for the parameters $b=0.1,b=1$ and $b=100$, respectively. Similarly for
the shock gravitational waves the curvature singularity occurs on the
hypersurface $\cos u\cdot\cosh u+\cos^{2}bv=1.$ The values of $u$ and $v$
confined to $0<u<\pi/2,$ $0<bv<\pi/2$ must be chosen in such a way to satisfy
$\cos u\cdot\cosh u+\cos^{2}bv<1.$ The Figures 3a, 3b and 3c illustrate the
shock wave plots for $b=0.1,b=1$ and $b=10$, respectively. Our study shows
that highly energetic em waves (i.e. greater $b$) udergo rotation shortly
after their encounter with the GW, while for the less energetic beams this
effect is delayed. Smooth variations, beside the local extrema is another
noticable effect in different plots. One fact, however, is evident that an em
wave reflecting from a cross polarized GW undergoes a Faraday rotation.
Further details of this effect can be obtained by studying more plots. Our
result applies also to the cosmic microwave background (CMB) radiation as well
as the radio astronomical fields. In other words, rotation in the polarization
vector in the CMB radiation can be attributed to the encounters with the
strong $(\times)$ moded GWs.

Considerations of mixed profile GWs and sandwich waves will be the subject of
a detailed analysis. Let us remark that encountering of an em wave with a
series of succesive impulsive GWs has already been considered \cite{GMO}.
Search for GWs through the Faraday rotation due to a test em wave in a GW
background has been considered by various authors \cite{RT},\cite{MC}.
Distinctly, our method takes the full non-linear effects into account by
employing the cross polarization mode of GWs.

\section{CONCLUSION}

In this paper, \ by using Einstein's theory of general relativity, we have
analysed the exact behavior of the polarization vector of a linearly polarized
em shock wave upon encountering with GWs. Expectedly, the Faraday's angle
emerges highly dependent on the type of the GW as well as the energy of the em
wave. It suggests that compared with the polarization changes  expected from
the quantum process of photon - electron scattering in the early universe, or
test field approximations, significant contributions to the polarization
vector can be imparted by the classical em-GW collisions.

\bigskip\newpage

\begin{center}
FIGURE CAPTION
\end{center}

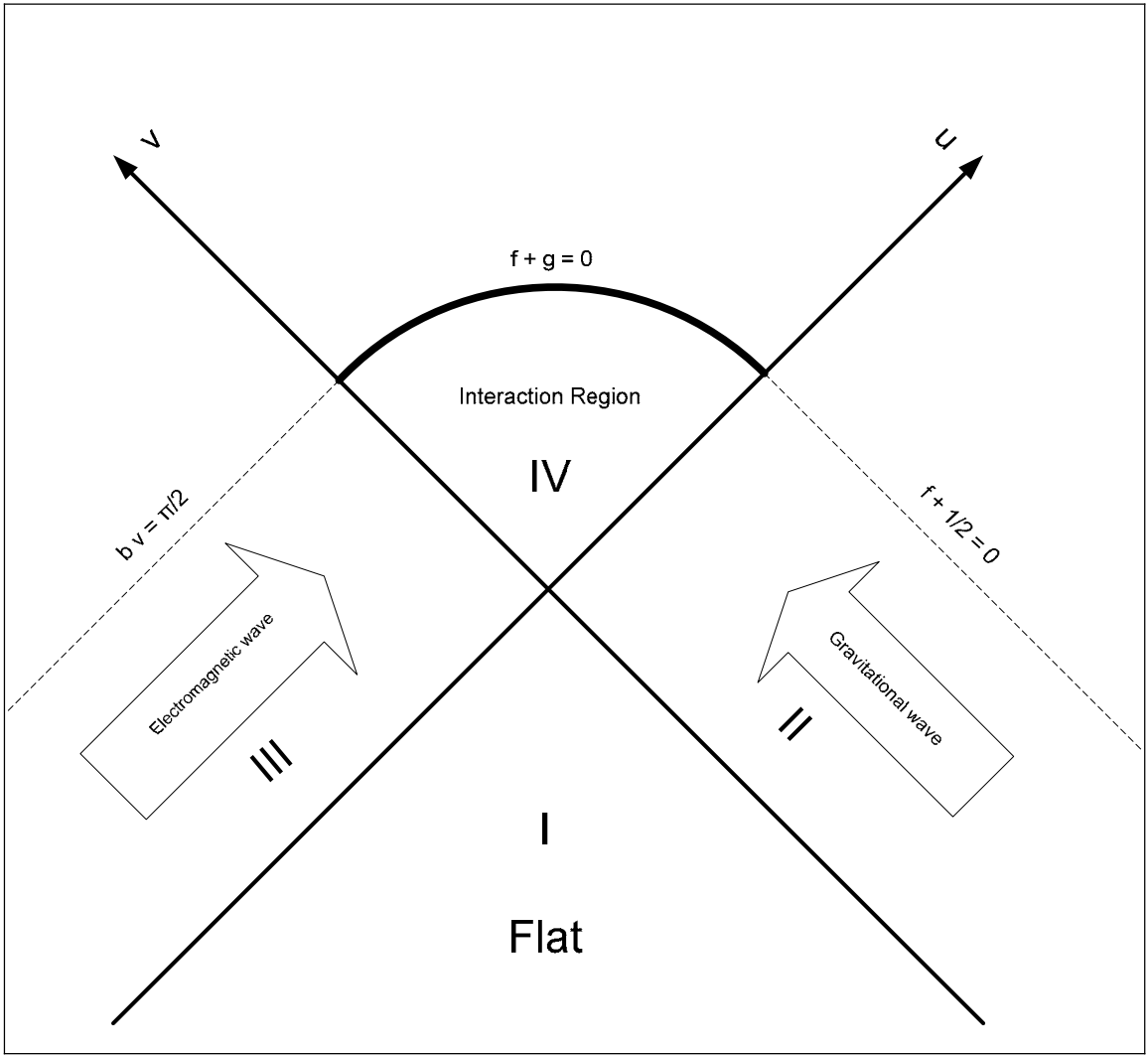
Figure 1.: The spacetime diagram describes the collision of plane em wave
propagating in one of the incoming region and plane GW propagating in the
other incoming region. The singular hypersurface occurs when $f+g=0$ in the
interaction region.

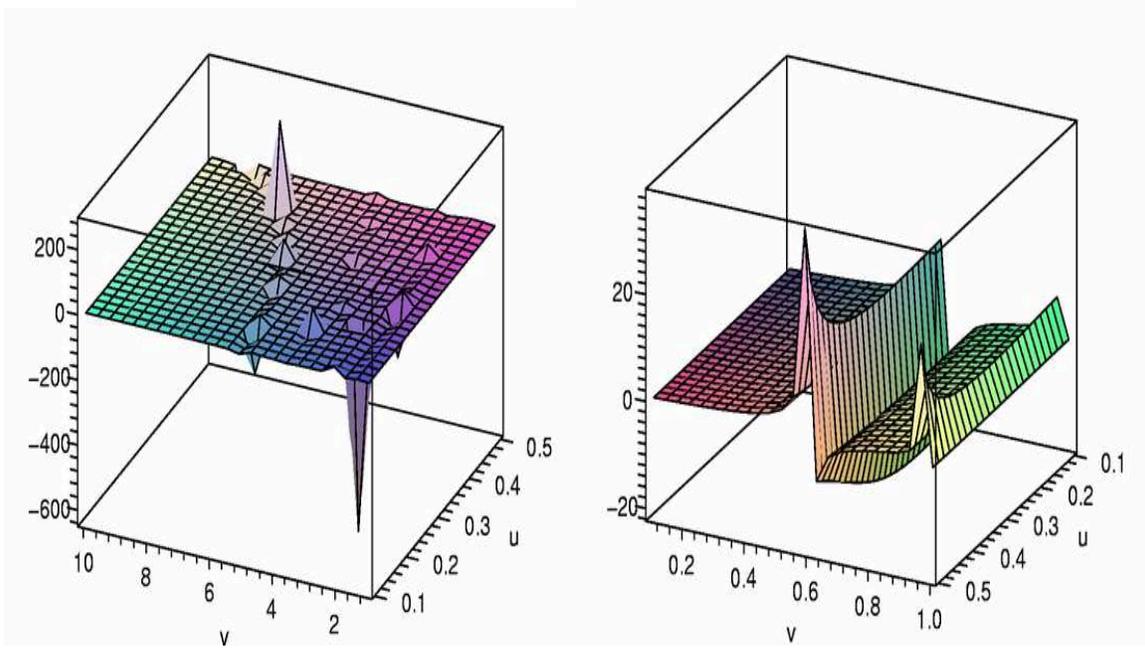
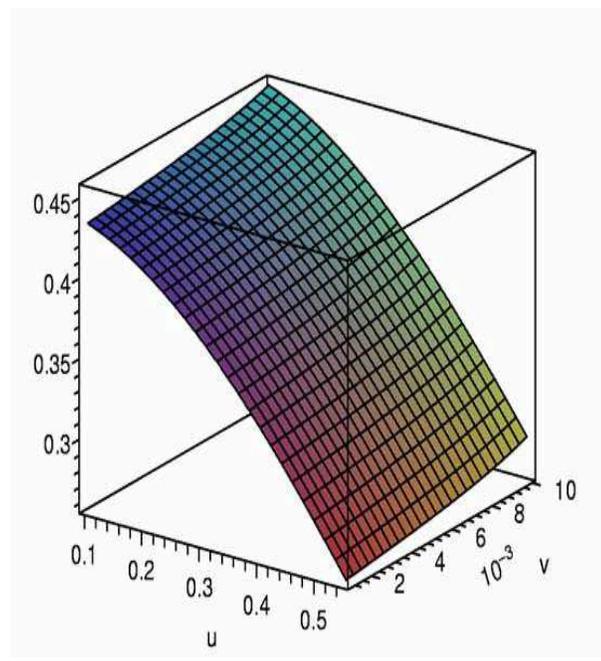
Figure 2. : The $\tan\theta$ plot when the electromagnetic shock wave
encounters a gravitational plane impulsive wave with the ($\times$)
polarization angle for $\alpha=45^{\circ}$, for frequencies; (a) $b=0.1$, (b)
$b=1$ and (c) $b=100$.

\bigskip

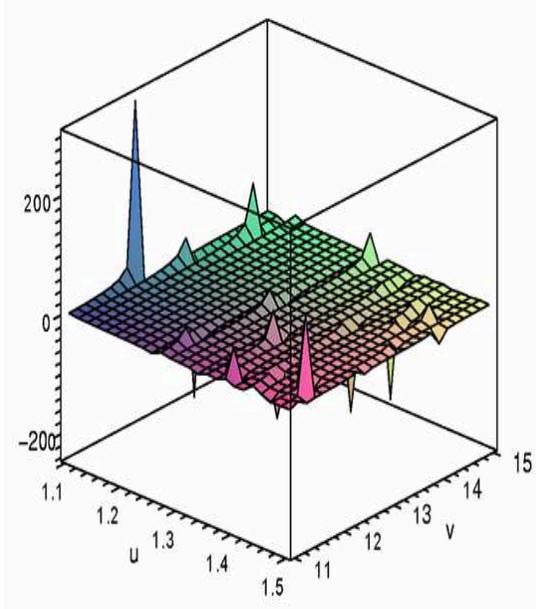
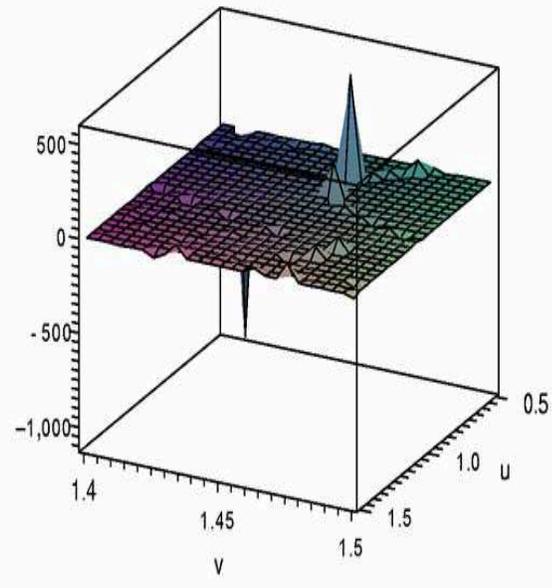
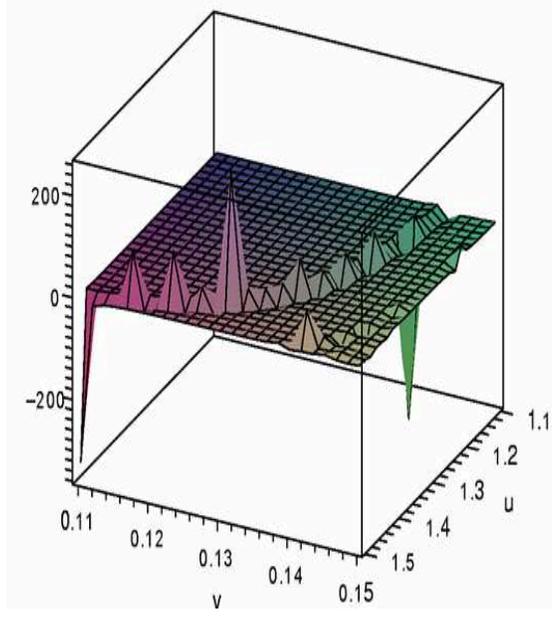
Figure 3. : The $\tan\theta$ plot when the electromagnetic shock wave
encounters a gravitational shock wave with the ($\times$) polarization angle
for $\alpha=45^{\circ}$, for frequencies; (a) $b=0.1$, (b) $b=1$ and (c)
$b=10$.

\end{document}